# *In situ* electric-field control of ferromagnetic resonance in the low-loss organic-based ferrimagnet V[TCNE]$_{x\sim2}$

Seth W. Kurfman[1], Andrew Franson[1], Piyush Shah[2], Yueguang Shi[3], Hil Fung Harry Cheung[4], Katherine E. Nygren[5], Mitchell Swyt[5], Kristen S. Buchanan[5], Gregory D. Fuchs[4], Michael E. Flatté[3,6], Gopalan Srinivasan[2], Michael Page[2], and Ezekiel Johnston-Halperin[†1]

[1]Department of Physics, The Ohio State University
[2]Materials and Manufacturing Directorate, Air Force Research Laboratory
[3]Department of Physics and Astronomy, University of Iowa
[4]Department of Physics, Cornell University
[5]Department of Physics, Colorado State University
[6]Department of Applied Physics, Eindhoven University of Technology

†Corresponding author email: johnston-halperin.1@osu.edu

**We demonstrate indirect electric-field control of ferromagnetic resonance (FMR) in devices that integrate the low-loss, molecule-based, room-temperature ferrimagnet vanadium tetracyanoethylene (V[TCNE]$_{x\sim2}$) mechanically coupled to PMN-PT piezoelectric transducers. Upon straining the V[TCNE]$_x$ films, the FMR frequency is tuned by more than 6 times the resonant linewidth with no change in Gilbert damping for samples with $\alpha = 6.5 \times 10^{-5}$. We show this tuning effect is due to a strain-dependent magnetic anisotropy in the films and find the magnetoelastic coefficient $|\lambda_s| \sim (1-4.4)$ ppm, backed by theoretical predictions from DFT calculations and magnetoelastic theory. Noting the rapidly expanding application space for strain-tuned FMR, we define a new metric for magnetostrictive materials, *magnetostrictive agility,* given by the ratio of the magnetoelastic coefficient to the FMR linewidth. This agility allows for a direct comparison between magnetostrictive materials in terms of their comparative efficacy for magnetoelectric applications requiring ultra-low loss magnetic resonance modulated by strain. With this metric, we show V[TCNE]$_x$ is competitive with other magnetostrictive materials including YIG and Terfenol-D. This combination of ultra-narrow linewidth and magnetostriction in a system that can be directly integrated into functional devices without requiring heterogeneous integration in a thin-film geometry promises unprecedented functionality for electric-field tuned microwave devices ranging from low-power, compact filters and circulators to emerging applications in quantum information science and technology.**

# Introduction

The use of electric fields for control of magnetism has been a long-term goal of magnetoelectronics in its many manifestations ranging from metal and semiconductor spintronics [1, 2], to microwave electronics [3, 4], to emerging applications in quantum information [5]. This interest arises from the potential for clear improvements in scaling, high-speed control, and multifunctional integration. However, while the promise of this approach is well established its realization has proven challenging due to the strong materials constraints imposed by the existing library of magnetic materials [3, 4]. The most common approach to achieving this local control is through linking piezoelectricity with magnetostriction to achieve electric-field control of magnetic anisotropy, either intrinsically through inherent coupling in multiferroic materials or extrinsically through piezoelectric/magnetic heterostructures [3]. Ideally, magnetic materials chosen for such applications should exhibit large magnetostriction, low magnetic damping and narrow linewidth (high-$Q$) ferromagnetic resonance (FMR), and robust mechanical stability upon strain cycling [4]. While the use of multiferroic materials promises relative simplicity in device design, they typically suffer from poor magnetic properties and minimal tunability. The alternative approach of employing heterostructures of magnetic thin films and piezoelectric substrates effectively creates a synthetic multiferroic exploiting the converse magnetoelectric effect (CME) and in principle allows independent optimization of piezoelectric and magnetic properties [6, 7, 8, 9, 10, 11].

Further, for applications that rely on magnetic resonance (*e.g.*, microwave electronics and magnon-based quantum information systems), the traditional metrics of magnetostriction, $\lambda_s$, or the CME coefficient, $A$, do not capture the critical parameters governing damping and loss in this regime (*e.g.*, linewidth or Gilbert damping coefficient). To date, materials with large magnetoelastic constants and CME coefficients (*e.g.,* Terfenol-D) suffer from high damping, broad magnetic resonance features, and are particularly fragile and brittle [4, 12]. Ferrites such as yttrium iron garnet (YIG), on the other hand, are attractive due to their low-loss magnetic resonance properties but typically exhibit minimal to moderate magnetoelastic coefficients. Further, these low-loss ferrites require high growth temperatures (800-900 °C) and lattice-matched substrates to produce high-quality material, which makes integrating these materials on-chip while maintaining low-loss properties challenging [13, 14, 15], and limits their applicability for magnetic microelectronic integrated circuits (MMIC). Accordingly, alternative low-loss, magnetostrictive materials with facile integration capabilities are desired for applications in electrically-controlled devices. Recently, a complementary material to YIG, vanadium tetracyanoethylene (V[TCNE]$_x$, $x \sim 2$), has gained significant interest from the spintronics and quantum information science and engineering (QISE) communities due to its ultra-low damping under magnetic resonance and benign deposition characteristics [16, 17, 18, 19, 20, 21, 22, 23, 24, 25, 26, 27].

Here we present the first systematic experimental study of the magnetostrictive properties of V[TCNE]$_x$. Composite heterostructures of V[TCNE]$_x$ films and piezoelectric substrates demonstrate shifts in the FMR frequency by 35 – 45.5 MHz, or more than 6 linewidths upon application of compressive strains up to $\varepsilon = -2.4 \times 10^{-4}$. Further, a systematic analysis shows that the Gilbert damping, $\alpha$, and inhomogeneous broadening linewidth, $\Gamma_0$, are insensitive to strain in this regime and robust to repeated cycling. Density-functional theory (DFT) calculations provide insight into the elastic and magnetoelastic properties of V[TCNE]$_x$ and predict a magnetoelastic coefficient $\tilde{\lambda}_{100}$ = −2.52 ppm. Experimental measures of the effective magnetoelastic constant, $\lambda_s$, determined by combining optical measurements of the distortion with the corresponding FMR frequency shifts, yield values of $\lambda_s$ = −(1 − 4.4) ppm, which is in good agreement with these DFT predictions. Finally, we define a new figure of merit, *magnetostrictive agility, $\zeta$*, as the ratio of the magnitude of the magnetoelastic coefficient to the FMR linewidth $\zeta = |\lambda_s|/\Gamma$ that more closely aligns with the performance requirements for emerging applications of magnetostrictive materials. These results establish a foundation for utilizing strain or acoustic (phononic) excitations for highly efficient strain-modulated magnetoelectronic devices based on V[TCNE]$_x$ and other next-generation organic-molecule-based magnetically-ordered materials for coherent information processing and straintronic applications [28].

# Results

V[TCNE]$_x$ is a room-temperature, organic-molecule-based ferrimagnet ($T_c \sim$ 600 K) that exhibits superb low-damping properties ($\alpha$ = (3.98 ± 0.22) × 10$^{-5}$) and high-quality factor FMR ($Q = f_R/\Gamma >$ 3,000) [17, 18, 19, 20, 21, 22, 23, 24, 25]. V[TCNE]$_x$ thin films are deposited via chemical vapor deposition (CVD) at relatively low temperature and high pressure (50 °C and 35 mTorr, respectively) and is largely insensitive to substrate lattice constant or surface termination [17, 20]. Further, V[TCNE]$_x$ can be patterned via *e*-beam lithography techniques without increase in its damping [19]. The highly coherent and ultra-low loss magnonic properties of V[TCNE]$_x$ have driven interest in applications in microwave electronics [26, 29] and magnon-based quantum information science and engineering (QISE) [21, 30, 31]. These benign deposition conditions, combined with patterning that does not degrade performance, highlight the versatility of V[TCNE]$_x$ for facile on-chip integration with pre-patterned microwave circuits and devices [26, 27, 29, 32, 33, 34]. These excellent magnetic properties are even more surprising given that V[TCNE]$_x$ lacks long-range structural order [25]. Early studies indicated that V[TCNE]$_x$ films do not exhibit magnetic anisotropy beyond shape effects due to this lack of long-range ordering [19, 20]. However, recent FMR studies on V[TCNE]$_x$ nanowires, microstructures, and thin films [19, 20, 21], coupled with combined DFT and electron energy loss spectroscopy (EELS) of the crystal structure [25], suggest there is a residual nematic ordering of the *c*-axis of the V[TCNE]$_x$ unit cell, giving rise to an averaged crystal field anisotropy that is sensitive to structural and thermally induced strain. However, dynamic measurements of these crystal fields and their dependence on strain are currently lacking, preventing a quantitative analysis of the magnetostrictive properties of this material.

For this work, PMN-PT/epoxy/V[TCNE]$_x$/glass heterostructure devices are fabricated such that upon electrically biasing the PMN-PT[001] substrate, the piezoelectric effect produces a lateral in-plane strain in the V[TCNE]$_x$ thin film, schematically shown in Fig. 1a. PMN-PT is selected for its strong piezoelectric effects and high strain coefficients ($d_{31} \sim -(500$ to $1{,}000)$ pm/V) to maximize the strain in the devices [3], and the epoxy encapsulation layer is selected to allow for device operation under ambient conditions [22]. This device structure allows investigation of the magnetoelastic properties of V[TCNE]$_x$ via standard FMR characterization and analysis. In the main text of this work, measurements on three devices denoted Samples 1 - 3 are presented. Sample 1 is measured via broadband FMR (BFMR) techniques. Sample 2 is studied via *X*-band (~9.8 GHz) cavity FMR techniques. Sample 3 is used to directly measure and calibrate strain in the devices via optical techniques. Additional devices characterized via BFMR techniques are presented in the Supplemental Information and their characteristics summarized in Table 1.

The BFMR response of Sample 1 is described in Fig. 1(b), where individual scans (inset) are fit to a Lorentzian lineshape to extract the resonance frequency as a function of applied field, $H_R$ vs $f_R$. This data can be modeled by considering the V[TCNE]$_x$ thin film as an infinite sheet with attendant shape anisotropy and with a uniaxial crystal field anisotropy oriented in the out-of-plane direction (as described above). Accordingly, the Kittel equation for ferromagnetic resonance reduces to [19, 20, 21]

$$f_R = \frac{\gamma}{2\pi}\sqrt{(H_R - H_{eff}\cos^2(\theta))(H_R - H_{eff}\cos(2\theta))} \qquad (1)$$

where $f_R = \omega/2\pi$ is the FMR resonance frequency, $\gamma$ is the gyromagnetic ratio, $H_R$ is the applied magnetic field at resonance, $H_{eff} = 4\pi M_{eff} = 4\pi M_s - H_\perp$ is the effective magnetization of the V[TCNE]$_x$ film with saturation magnetization $4\pi M_s$ and uniaxial strain-dependent anisotropy field $H_\perp$, and $\theta$ describes the orientation of the external field as defined in Fig. 1(a). This equation is valid for films where $H_R \gg 4\pi M_{eff}$, and is appropriate here as the effective magnetization for V[TCNE]$_x$ is typically ~100 G [19] while the resonance field is typically between 3500 – 3650 G at *X*-band frequencies (9.86 GHz) for all magnetic field orientations. As-grown films exhibit no in-plane anisotropy, consistent with the literature [19, 20, 21], and so $\phi$-dependences are neglected. When the external magnetic field is held out-of-plane ($\theta = 0°$), Eq. (1) reduces to

$$f_R = \frac{\gamma}{2\pi}(H_R - 4\pi M_{eff}) \qquad (2)$$

Further information about the magnetic damping in thin films can be revealed by comparing the FMR full-width-half-max (FWHM) frequency linewidth, $\Gamma$, to the FMR resonance frequency, $f_R$, via (for $4\pi M_{eff} \ll H_R$ and $\Gamma \ll f_R$ [19])

$$\Gamma = 2\alpha f_R + \Gamma_0 \qquad (3)$$

where $\alpha$ is the (dimensionless) Gilbert damping constant and $\Gamma_0$ is the inhomogeneous broadening. It should be noted this form for the Gilbert damping utilizing the frequency-swept linewidth is appropriate directly for out-of-plane magnetized thin films due to symmetry conditions resulting in the linear relationship between $H_R$ and $f_R$ [19]. Accordingly, Eqs. (1 - 3) show that performing FMR at various frequencies, fields, and magnetization orientations with and without applied strains in V[TCNE]$_x$ thin films should provide information regarding the magnetoelastic properties of V[TCNE]$_x$.

Fitting the data from Sample 1 to Eq. (2) reveals an effective magnetization $4\pi M_{eff} = 106.2$ G and gyromagnetic ratio $|\gamma|/2\pi = 2.756$ MHz/Oe, consistent with literature [17, 18, 19, 20, 21, 22]. The Lorentzian fits of the FMR response also reveal the linewidth $\Gamma$ as a function of the resonant frequency, seen in Fig. 1(c), where fitting to Eq. (3) yields $\alpha = 1.02 \pm 0.52 \times 10^{-4}$ and $\Gamma_0 = 8.48 \pm 1.22$ MHz in Sample 1. These damping characteristics are also consistent with literature values for V[TCNE]$_x$ [19, 27], and show that the devices incorporate high-quality magnetic films exhibiting superb low-damping properties [32, 33, 34].

Moving beyond measurements of the as-grown strain-free sample, the FMR response of the device is now measured while straining the V[TCNE]$_x$ film (Fig. 1(d)). Comparing the FMR response of Sample 1 with no applied strain ($E_B = 0$ kV/cm) and maximum-applied strain ($E_B = 13.3$ kV/cm) yields a shift in the resonance frequency of 45.5 MHz at a resonance frequency $f_R = 9.8$ GHz, corresponding to a CME coefficient $A = 3.38$ MHz cm/kV (1.23 Oe cm/kV). It is worth noting that while this absolute shift in frequency, and consequent value for CME, is modest when compared to other magnetostrictive materials [4], it represents a shift of over 6 magnetic resonance linewidths due to the ultra-low damping and narrow FMR linewidths of the V[TCNE]$_x$ thin film. This ability to shift cleanly on and off resonance with an applied electric field is central to the functionality of many dynamically tuned MMIC devices, motivating a more in depth and systematic investigation of this phenomenon.

The magnetostriction in this composite device is explored by biasing the piezoelectric transducer between 0 kV/cm and 13.3 kV/cm, and the shift in the resonance frequency (for $\theta = 0°$) tracks the linear strain produced by the transducer [35], as seen in Fig. 2(a). For maximally strained films, fitting to Eq. (2) now reveals $4\pi M_{eff} = 122.9$ G, a difference of +16.7 G between $E_B = 0$ kV/cm and $E_B = 13.3$ kV/cm (14% change). Panels (b-d) of Fig. 2 show the FMR linewidth $\Gamma$, inhomogeneous broadening $\Gamma_0$, and Gilbert damping $\alpha$, for Sample 1 as a function of applied electric field (strain). These parameters do not vary over the entire tuning range and are robust to repeated cycling (>300 cycles - see Supplemental Information).

This stability indicates that the shift in resonance frequency is due to a true magnetoelastic effect under linear deformation rather than some fatigue induced structure or morphology change in the film, and further demonstrates the potential for device applications. Finally, it is noteworthy that the linewidths and damping coefficients observed in these proof of principle devices are much narrower than typical magnetostrictive materials, but are roughly twice the value observed in optimized bare V[TCNE]$_x$ films ($\Gamma$ is typically ~3 MHz [17, 19, 25]). This suggests that the tuning ratio of 6 times the linewidth may be further extended to more than 10 times the linewidth in fully optimized devices [19, 25].

To confirm that these shifts in the resonance position are due to strain-dependent crystal-field anisotropy in V[TCNE]$_x$ as prior studies suggest [20, 21], angular-dependent measurements on unstrained and maximally strained films are performed. Sample 2 is mounted in an $X$-band (~9.8 GHz) microwave cavity so that the structure can be rotated to vary the polar angle, $\theta$. In-plane ($\theta = 90°$) and out-of-plane ($\theta = 0°$) FMR spectra are shown in Supplemental Fig. S1, with FWHM linewidths of 2.17 Oe (5.97 MHz) and 2.70 Oe (7.45 MHz), respectively. By tracking the resonance field as a function of rotation and fitting to Eq. (1) the effective magnetization $H_{eff}= 4\pi M_{eff}= 74.0$ G is extracted for Sample 2, as seen in Fig. 3. The difference in $4\pi M_{eff}$ between Samples 1 and 2 can be attributed to sample-to-sample variation and remains consistent with literature values [19]. Repeating the measurement with an applied bias of 13.3 kV/cm to the PMN-PT reveals an increase of $4\pi M_{eff}$ to 79.4 G, an increase of 5.4 Oe (8% change), which is like the change observed in Sample 1. This confirms that strain is modulating the magnetic anisotropy in V[TCNE]$_x$ through the crystal field term $H_\perp$ where $4\pi M_{eff}= 4\pi M_s − H_\perp$. This strain-dependent crystal field $H_\perp$ is consistent with and supports previous measurements of V[TCNE]$_x$ with both thermally and structurally induced strain [19, 20, 21].

An approximate upper bound to the strain in these devices can be simply calculated through the relation $\varepsilon = d_{31}E_B = (d_{31}V_B)/t \sim -(6 - 12) \times 10^{-4}$ for typical PMN-PT $d_{31}$ piezo coefficients [4, 35]. However, the addition of epoxy, V[TCNE]$_x$, and the glass substrate affect the overall stiffness of the device, thereby the piezo coefficient changes from $d_{31}$ of the bare piezo to an effective coefficient $d_{eff}$ of the entire stack. This $d_{eff}$ is directly measured by exploiting the color change of V[TCNE]$_x$ upon laser heating [23] to pattern fiducial marks on the samples and monitor their positions under strain using optical microscopy (see Supplemental Information). This approach yields an effective piezoelectric coefficient of $d_{eff} \sim -180$ pm/V or strain of $\varepsilon \sim -2.4 \times 10^{-4}$, reasonable for the PMN-PT heterostructures used here [4, 35].

Density functional theory calculations on the relaxed and strained V[TCNE]$_x$ unit cell provide further insight into the elastic and magnetoelastic properties of V[TCNE]$_x$. These properties are calculated using the Vienna ab initio Simulation Package (VASP) (version 5.4.4) with a plane-wave basis, projector-augmented-wave pseudopotentials [36, 37, 38, 39], and hybrid functional treatment of Heyd-Scuseria-Ernzerhof (HSE06) [40, 41]. The experimentally verified [25] local structure of the V[TCNE]$_x$ unit cell is found by arranging the central V atom and octahedrally-coordinated TCNE ligands according to experimental indications [42, 43, 44, 45, 46], and subsequently allowing the structure to relax by minimizing the energy. These DFT results previously produced detailed predictions of the structural ordering of V[TCNE]$_x$, along with the optoelectronic and inter-atomic vibrational properties of V[TCNE]$_x$ verified directly by EELS [25] and Raman spectroscopy [23], respectively. This robust and verified model therefore promises reliable insight into the elastic and magnetoelastic properties of V[TCNE]$_x$.

The magnetoelastic energy density for a cubic lattice $f = f_{el} + f_{me} = E/V$ is a combination of the elastic energy density

$$f_{el}^{cubic} = \frac{1}{2}C_{11}\left(\varepsilon_{xy}^2 + \varepsilon_{yz}^2 + \varepsilon_{zx}^2\right) + 2C_{44}\left(\varepsilon_{xx}^2 + \varepsilon_{xx}^2 + \varepsilon_{xx}^2\right) + C_{12}\left(\varepsilon_{xx}\varepsilon_{yy} + \varepsilon_{yy}\varepsilon_{zz} + \varepsilon_{zz}\varepsilon_{xx}\right) \quad (4)$$

where $C_{ij}$ are the elements of the elasticity tensor and $\varepsilon_{ij}$ are the strains applied to the cubic lattice, and the magnetoelastic coupling energy density

$$f_{me}^{cubic} = B_0\left(\varepsilon_{xy} + \varepsilon_{yz} + \varepsilon_{zx}\right) + B_1\left(\alpha_x^2\varepsilon_{xx} + \alpha_y^2\varepsilon_{xx} + \alpha_z^2\varepsilon_{xx}\right) + 2B_2\left(\alpha_x\alpha_y\varepsilon_{xy} + \alpha_y\alpha_z\varepsilon_{yz} + \alpha_z\alpha_x\varepsilon_{zx}\right) \quad (5)$$

where $B_i$ are the magnetoelastic coupling constants and $\alpha_i$ where $i \in \{x,y,z\}$ represent the cosines of the magnetization vector [47].

The elastic tensor $C = C_{ij}$ for V[TCNE]$_x$ is found by applying various strains to the unit cell and observing the change in the energy. The calculated $C_{ij}$ tensor results in a predicted Young's modulus for V[TCNE]$_x$ $Y_V$ = 59.92 GPa. By directly applying compressive and tensile in-plane strains to the DFT unit cell (i.e. in the *equatorial* TCNE ligand plane [25]), one may calculate the overall change in the total energy density, both parallel and perpendicular to the easy axis. The difference between these two, $\Delta E$, is the magnetic energy density change, which is proportional to the magnetoelastic coupling constant $B_1$ [47]

$$\Delta E/V = -(\nu_{2D} + 1)B_1\varepsilon_\parallel \quad (6)$$

where $\nu_{2D}$ is the 2-dimension in-plane Poisson ratio and $\varepsilon_\parallel$ is the applied in-plane (equatorial TCNE plane) epitaxial strain while allowing out-of-plane (apical TCNE direction) relaxation. The elastic and magnetoelastic coefficients are related via the magnetostriction constant $\lambda_{100} = \lambda_s$ via

$$\lambda_{100} = -\frac{2}{3}\frac{B_1}{C_{11} - C_{12}}. \quad (7)$$

As a result, the calculated changes of the magnetoelastic energy density with strain provide direct predictions of the elasticity tensor ($C_{ij}$) and magnetoelastic coefficients ($B_i$) for V[TCNE]$_x$. For polycrystalline samples of cubic materials, the overall (averaged) magnetoelastic coefficient $\lambda_s$ also considers the off-axis contribution from $\lambda_{111}$ such that $\lambda_s = (2/5)\lambda_{100} + (3/5)\lambda_{111}$. However, the off-axis component is not considered here for two reasons: (i) the apical TCNE ligands are assumed to align along the out-of-plane direction ($z$-axis, $\theta = 0°$), and (ii) difficulties in calculating the magnetoelastic energy density changes upon applying a shear strain that provides the estimate of $B_2$ needed to calculate $\lambda_{111}$. The former argument is reasonable as previous experimental results indicate the magnetocrystalline anisotropy from strain is out-of-plane [21], consistent with the ligand crystal field splitting between the equatorial and apical TCNE ligands [25]. Further, the lack of in-plane ($\phi$-dependent) anisotropy suggests the distribution in the plane averages out to zero. Therefore, the magnetoelastic coefficient calculated here considers an average of the in-plane $C_{ii}$ components in determining $\lambda_{100}$. That is, the DFT predicts a magnetoelastic coefficient for V[TCNE]$_x$ of

$$\lambda_s = -\frac{2}{3}\frac{B_1}{\overline{C}_{IP} - C_{12}} \tag{8}$$

where $\overline{C}_{IP} = (1/2)(C_{11} + C_{22}) = 60.56$ GPa and $C_{12} = 37.84$ GPa. Accordingly, utilizing the calculated value of $B_1 = 85.85$ kPa (see Supplemental Information) predicts a theoretically calculated $\tilde{\lambda}_{100} = -2.52$ ppm for V[TCNE]$_x$ magnetized along the apical TCNE ligand (i.e. $\theta = 0°$).

Combining these ferromagnetic resonance, direct strain measurements, and DFT calculations provides the information necessary to determine the magnetoelastic properties of V[TCNE]$_x$. Here, we follow the convention in the literature using the magnetoelastic free energy form from the applied stress $\sigma = Y\varepsilon$ to the magnetostrictive material, $F_{me} = (3/2)\lambda_s\sigma$ [2, 3, 4]. Accordingly, this free energy yields an expression for the strain-dependent perpendicular (out-of-plane) crystal field [3]

$$H_\perp = \frac{12\pi\lambda_S Y}{4\pi M_s} d_{31} E_B \tag{9}$$

where $\lambda_S$ is the magnetoelastic coefficient, $Y$ is the Young's modulus of the magnetic material, $d_{31}$ is the piezoelectric coefficient of the (multiferroic) crystal, and $E_B$ is the electric field bias. Here, $\varepsilon = d_{31}E_B$ is the strain in the magnetic layer obtained based on the assumption that the electrically induced strain is perfectly transferred to the magnetic film. For this study, the direct optical measurement of the strain in the V[TCNE]$_x$ films allows the modification of Eq. 9 by replacing $d_{31}E_B$ by the measured $\varepsilon = d_{eff}E_B = -2.4 \times 10^{-4}$ to account for the mechanical complexity of the multilayered device. The magnetoelastic coefficient of V[TCNE]$_x$ can then be calculated from Eq. 9 using the values of $4\pi M_S$ and $H_\perp$ from FMR characterization, the direct measurement of $\varepsilon$ from optical measurements, and the calculated value of $Y_V = 59.92$ GPa from DFT. Accordingly, inserting the corresponding values into Eq. 9 yields a magnetoelastic constant for V[TCNE]$_x$ of $\lambda_s \sim -1$ ppm to $\lambda_s \sim -4$ ppm for the devices measured here. This range shows excellent agreement with the DFT calculations of the magnetoelastic coefficient $\tilde{\lambda}_{100} = -2.52$ ppm from Eq. 8. This agreement provides additional support for the robustness of the DFT model developed in previous work [23, 25].

Further, comparison with past studies of the temperature dependence of $H_{eff}$ [21] allows for the extraction of the thermal expansion coefficient of V[TCNE] $_x$, $\alpha_{th}$ = 11 ppm/K, at room temperature.

## Discussion

The results above compare V[TCNE]$_x$ thin films to other candidate magnetostrictive materials using the established metrics of CME and $\lambda_s$. However, while these parameters are effective in capturing the impact of magnetoelastic tuning on the DC magnetic properties of magnetic thin films and magnetoelectric devices, they fail to capture the critical functionality for dynamic (AC) magnetoelectric applications: the ability to cleanly tune on and off magnetic resonance with an applied electric field. For example, Terfenol-D is considered a gold standard magnetostrictive material due to its record large magnetoelastic coefficient $\lambda_s$ up to 2,000 and CME coefficients $A$ as large as 590 Oe cm/kV [3]. However, due to its broad ~1 GHz FMR linewidths, large $4\pi M_{eff}$ > 9,000 G, high Gilbert damping $\alpha = 6 \times 10^{-2}$, and brittle mechanical nature, it is not practical for many applications in MMIC. As a result, we propose a new metric that appropriately quantifies the capability of magnetostrictive materials for applications in microwave magnonic systems [28, 33, 34] that takes into account both the magnetostrictive characteristics *and* the linewidth (loss) under magnetic resonance of a magnetically-ordered material. Accordingly, a *magnetostrictive agility* $\zeta$ is proposed here, which is the ratio of the magnetoelastic coefficient $\lambda_s$ to the FMR linewidth (in MHz) $\zeta(f_R)$ = $|\lambda_s|/\Gamma$. For the V[TCNE]$_x$ films studied here the magnetostrictive agility at *X*-band frequencies (9.8 GHz) is in the range $\zeta$ = {0.164 – 0.660}, comparable to YIG $\zeta$ = {0.139 − 0.455} and Terfenol-D $\zeta$ = {0.301 – 0.662} as shown in Table 1. Further, we note that the growth conditions under which high-quality V[TCNE]$_x$ films can be obtained make on-chip integration with microwave devices significantly more practical than for YIG, and that the narrow linewidth (low loss) is more attractive for applications such as filters and microwave multiplexers than Terfenol-D.

## Conclusion

We have systematically explored indirect electric-field control of ferromagnetic resonance in the low-loss organic-based ferrimagnet V[TCNE]$_x$ in V[TCNE]$_x$/PMN-PT heterostructures. These devices demonstrate the ability to shift the magnetic resonance frequency of V[TCNE]$_x$ by more than 6 linewidths upon application of compressive in-plane strains $\varepsilon \sim 10^{-4}$. Further, we find there is no change in the magnetic damping of the films with strain and that the samples are robust to repeated cycling (> 300 cycles), demonstrating the potential for applications in MMIC without sacrificing the ultra-low damping of magnetic resonance in V[TCNE]$_x$. The changes in the FMR characteristics along with direct optical measurements of strain provide an experimentally determined range for the magnetoelastic coefficient, $\lambda_S$ = −(1 − 4.3) ppm, showing excellent agreement to DFT calculations of the elastic and magnetoelastic properties of V[TCNE]$_x$. Finally, we present a discussion on the metrics used in the magnetostriction community wherein we point out the shortcomings on the commonly used metrics of the magnetostriction and CME coefficients. In this context, we propose a new metric, the magnetostrictive agility, $\zeta$, for use of magnetoelastic materials for coherent magnonics applications.

These results develop the framework necessary for extended studies into strain-modulated magnonics in V[TCNE]$_x$. Additionally, these magnetoelastic properties in V[TCNE]$_x$ suggest that large phonon-magnon coupling in V[TCNE]$_x$ might be achieved, necessary and useful for applications in acoustically-driven FMR (ADFMR) or, in conjunction with high-Q phonons, for quantum information applications [28]. Other recent work has identified V[TCNE]$_x$ as a promising candidate for QISE applications utilizing superconducting resonators [31] and NV centers in diamond ranging from enhanced electric-field sensing [5] to coupling NV centers over micron length scales [30]. These findings lay a potential framework for investigating the utilization of V[TCNE]$_x$ in quantum systems based on magnons and phonons.

# Acknowledgments

S. W. K. developed the project idea, and S. W. K., E. J.-H., P. S., and M. P. developed the project plan for experimental analysis. S. W. K. fabricated the V[TCNE]$_x$ heterostructure devices, performed FMR characterization and analysis, and wrote the manuscript. A. F. developed the analysis software used for fitting FMR linewidths and extracting parameter fits. P. S., G. S., and M. P. provided PMN-PT substrates. Y. S. performed and analyzed DFT calculations of the elasticity tensor and the magnetoelastic coefficients for V[TCNE]$_x$. H. F. H. C. performed and analyzed optical measurements of strain in the devices. K. E. N. and M. S. performed BLS measurements on V[TCNE]$_x$/Epoxy devices to extract elastic properties. All authors discussed the results and revised the manuscript. S. W. K., A. F., and E. J.-H. were supported by NSF DMR-1808704. P. S., G. S, and M. P. were supported by the Air Force Office of Scientific Research (AFOSR) Award No. FA955023RXCOR001. The research at Oakland University was supported by grants from the National Science Foundation (DMR-1808892, ECCS-1923732) and the Air Force Office of Scientific Research (AFOSR) Award No. FA9550-20-1-0114. Y. S. and M. E. F. were supported by NSF DMR-1808742. H. F. H. C. and G. D. F. were supported by the DOE Office of Science (Basic Energy Sciences) grant DE-SC0019250. K. E. N., M. S., and K. S. B. were supported by NSF-EFRI grant NSF EFMA-1741666. The authors thank and acknowledge Georg Schmidt, Hans Hübl, and Mathias Kläui for fruitful discussions.

# Methods

### Synthesis of V[TCNE]$_x$ and Device Fabrication.

V[TCNE]$_x$ films are deposited via ambient-condition chemical vapor deposition (CVD) in a custom CVD reactor inside an argon glovebox ($O_2$ < 1 ppm, $H_2O$ < 1 ppm) in accordance with literature [17, 18, 19, 20, 21, 22, 23, 24, 25, 26, 27, 29, 31]. Argon gas flows over TCNE and V(CO)$_6$ precursors that react to form a V[TCNE]$_x$ thin film on the substrates. The pressure inside the CVD reactor for all growths was 35 mmHg, and TCNE, V(CO)$_6$, and the substrates are held at 65°C, 10°C, and 50°C, respectively. All substrates were cleaned via solvent chain (acetone, methanol, isopropanol, and deionized (DI) water (×2)) and dried with $N_2$, followed by a 10 minute UV/Ozone clean in a UVOCS T10x 10/OES to remove any residual organic contaminants.

Nominally 400 nm V[TCNE]$_x$ films are deposited onto microscope cover glass substrates ($t = 100\mu$m). These V[TCNE]$_x$/glass substrates are then mechanically fixed to a PMN-PT transducer (4 mm×10 mm ×0.15 mm) with an OLED epoxy (Ossila E130) to create a PMN-PT/Epoxy/V[TCNE]$_x$/Glass heterostructure. The epoxy here not only protects V[TCNE]$_x$ from oxidation [22], but also propagates lateral strain into V[TCNE]$_x$ film from the piezo transducer upon biasing. While the primary deformation in the piezo transducer is along the poling direction of the PMN-PT ($z$), the distortion of the PMN-PT in the thickness direction also produces a lateral in-plane strain in the PMN-PT through the Poisson effect (i.e. one must consider here the $d_{31}$ piezo coefficient of PMN-PT). Therefore, the primary strain experienced by the V[TCNE]$_x$ film is in-plane. The PMN-PT electrodes are connected to a Keithley 2400 voltage source so that electric fields up to $E_B = V_B/t_{PMN-PT} = 13.3$ kV/cm can be applied across the PMN-PT layer.

## Ferromagnetic Resonance Characterization

Broadband FMR (BFMR) measurements on Sample 1 and Supplemental Devices A-C were taken using a commercial microstrip (Southwest Microwave B4003-8M-50) and Agilent N5222A vector network analyzer (VNA). The devices are mounted so that the magnetic field is normal to the V[TCNE]$_x$ film ($\theta = 0°$). S21 measurements (P = −20 dBm) show the FMR peak upon matched magnetic field and frequency conditions in accordance with Eq. 2. A Keithley 2400 Sourcemeter is used to apply up to 200 V to the piezoelectric transducers – accordingly, the maximum-applied strain in the 150 μm PMN-PT corresponds to an electric field $E_B = 13.3$ kV/cm as mentioned in the main text.

All angular-dependent FMR measurements (Sample 2) were performed in a Bruker *X*-band (~9.6 GHz) EPR (Elexsys 500) spectrometer. The frequency of the microwave source is tuned to match the resonant frequency of the cavity before each scan to ensure optimal cavity tuning. All scans had a 0.03 G modulation field at 100 kHz modulation frequency and were performed at the lowest possible microwave power (0.2 μW) to prevent sample heating and non-linear effects distorting the FMR lineshape. The V[TCNE]$_x$/PMN-PT devices are mounted on a sapphire wafer and loaded into glass tubes for FMR measurements such that the samples can be rotated in-plane (IP: $\theta = 90°$) to out-of-plane (OOP: $\theta = 0°$) for FMR measurements in 10 degree increments, where resonance occurs upon matched field and frequency conditions according to Eq. 1.

## Density Functional Theory Calculations

The pseudopotentials used are default options from VASP's official PAW potential set, with five valence electrons per vanadium, four per carbon and five per nitrogen [36, 37, 38, 39]. For the rest of the calculation we used 400 eV for the energy cutoff and a Γ centered 5x5x3 k-mesh sampling. From these results, the elastic tensor $C_{ij}$ for V[TCNE]$_x$ is calculated. Using the elastic tensor, the Young's modulus for V[TCNE]$_x$ is averaged over the $C_{11}$, $C_{22}$, and $C_{33}$ components to yield $Y_V = 59.92$ GPa. From the DFT calculations, the full elastic matrix from the $C_{ij}$ is given by (in units of GPa)

$$C_{ij} = \begin{bmatrix} 66.44 & 37.84 & 7.96 & 1.38 & -0.20 & 0.53 \\ 37.84 & 54.68 & 3.79 & 0.09 & -1.55 & -0.37 \\ 7.96 & 3.79 & 58.64 & -0.69 & 0.76 & 0.31 \\ 1.38 & 0.09 & -0.69 & 35.16 & 0.25 & -0.95 \\ -0.20 & -1.55 & 0.76 & 0.25 & 6.65 & -0.17 \\ 0.53 & -0.37 & 0.31 & -0.95 & -0.17 & 9.94 \end{bmatrix}$$

**Optical Measurements of Strain in V[TCNE]$_x$**

V[TCNE]$_x$ films can be patterned via laser heating techniques, whereupon the material changes color when heated above its thermal degradation temperature (∼ 370 K) [16, 23]. To more appropriately calibrate strain in the V[TCNE]$_x$ films versus applied bias, we directly measure the deformation in the films by exploiting the color change of V[TCNE]$_x$ upon laser heating [23] and optical microscopy techniques. Fresh V[TCNE]$_x$/PMN-PT devices are exposed to a focused laser spot to create *ad hoc* fiducial marks on the film in Sample 3 (Supplementary Fig. S5) in a 50 $\mu$m × 50 $\mu$m square. By measuring the distance between these laser-written structures with and without applied strain, we can precisely and directly measure the strain in the V[TCNE]$_x$ films upon electric bias thus allowing a more precise calculation of magnetoelastic coefficients. Using these methods, we apply a bias of 13.3 kV/cm on Sample 3 and find a strain $\varepsilon \sim 2.4 \times 10^{-4}$ which is in reasonable agreement with estimated values of strain using the thickness of the PMN-PT (150 $\mu$m) and typical piezo coefficient $d_{31} \sim 500 - 1000$ pm/V).

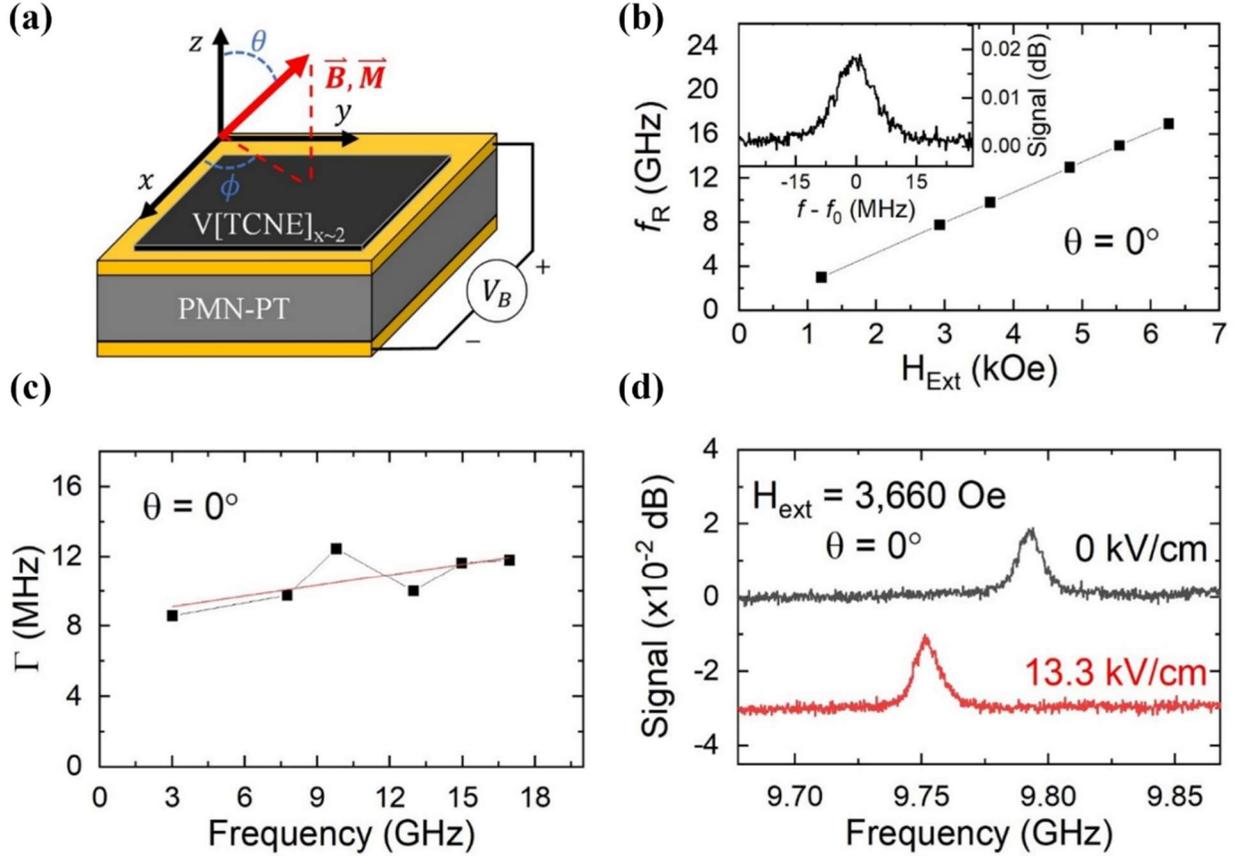

**Figure 1:** (a) Effective device schematic, coordinate system, and wiring diagram for V[TCNE]$_x$/PMN-PT heterostructures. (b) Ferromagnetic resonance frequency $f_R$ vs external field $H_{ext}$ with the field held OOP ($\theta = 0°$) measured via BFMR. The external field is held constant as the microwave frequency is swept. (Inset) Representative BFMR scan at $f_0 = 9.8$ GHz, $H_{ext} = 3,660$ Oe. (c) FMR linewidth $\Gamma$ versus FMR frequency for OOP field ($\theta = 0°$). A linear fit (red line) extracts the dimensionless Gilbert damping parameter $\alpha = 1.02 \pm 0.52 \times 10^{-4}$ and the inhomogeneous broadening $\Gamma_0 = 8.48 \pm 1.22$ MHz. (d) BFMR scans for unstrained (0 kV/cm – black) and maximally strained (13.3 kV/cm – red). The shift in the FMR frequency is ~45 MHz, a shift ~4 linewidths.

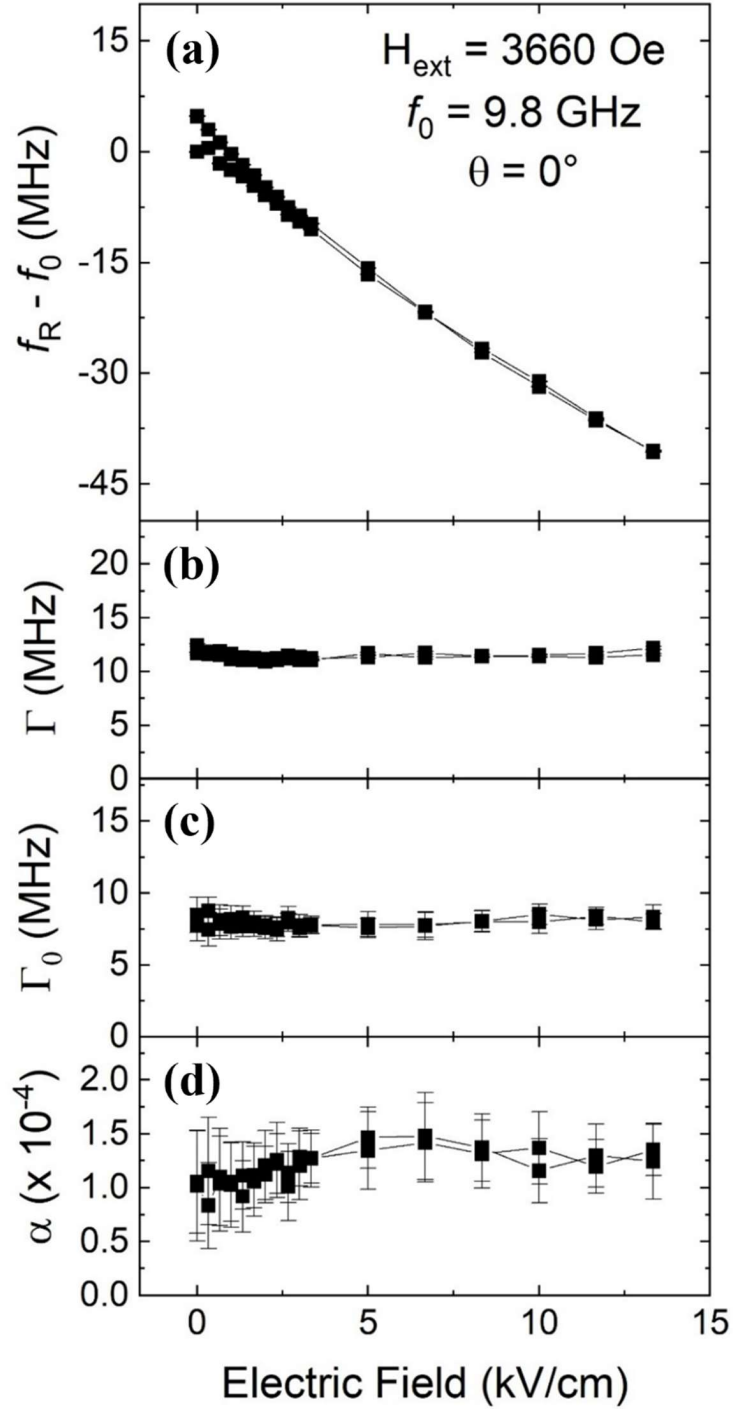

**Figure 2:** V[TCNE]$_x$ damping analysis with applied strain: (a) Plot showing differential (shifted) FMR resonance position $f_R-f_0$ where $f_0$ is the resonance at 9.8 GHz, (b) FWHM linewidth $\Gamma$, (c) inhomogeneous broadening $\Gamma_0$, and (d) Gilbert damping $\alpha$ versus applied electric field bias $E_B$. While the resonance position shifts by multiple linewidths, there is negligible effect in the linewidth or damping of the material.

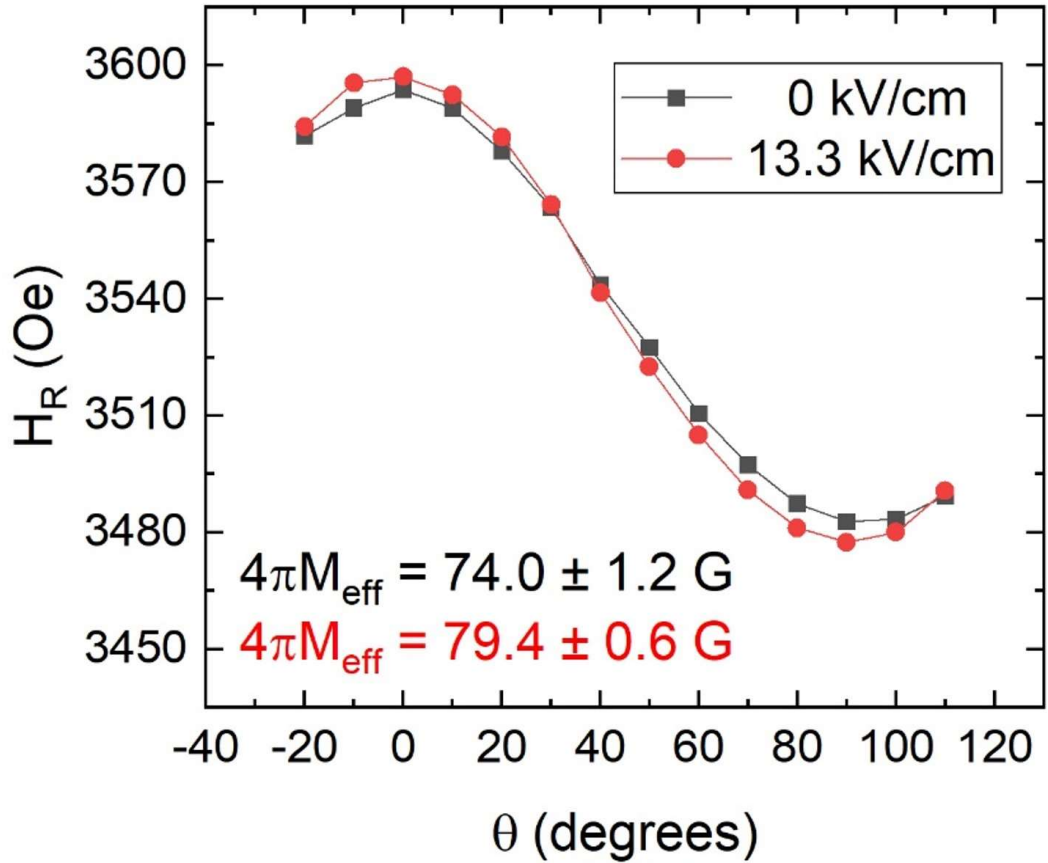

**Figure 3:** Cavity *X*-band FMR measurements on V[TCNE]$_x$/PMN-PT devices. Angular dependence of FMR resonant field H$_R$ at $f_R \sim$ 9.6 GHz is measured without (black squares) and with (red circles) strain. In-plane and out-of-plane peak-to-peak (FWHM) linewidths are 1.25 (2.16) Oe and 1.56 (2.70) Oe, respectively.

| Material | $4\pi M_{\text{eff}}$ (G) | $|\lambda_s|$ (ppm) | $\alpha$ | $\Gamma$ @ 9.8 GHz | $\zeta$ @ 9.8 GHz | Ref. |
|---|---|---|---|---|---|---|
| V[TCNE]$_x$ S1 | 106.2 | 4.356 | $1.02\times10^{-4}$ | 12 MHz | 0.363 | This work. |
| V[TCNE]$_x$ S2 | 74.0 | 0.978 | N/A | 2.16 Oe (5.96 MHz)* | 0.164 | This work. |
| V[TCNE]$_x$ Supp. Devices | ~100 | 1.34 - 2.97 | $6.5\times10^{-5}$ | 4.5 - 6.3 MHz | 0.213–0.66 | This work. |
| Thin Film YIG | 1750 | 1.4 - 4.6 | $6.15\times10^{-5}$ | 3.5 Oe (10.1 MHz)* | 0.139 - 0.455 | [10, 13] |
| Terfenol-D (Tb$_{0.3}$Dy$_{0.7}$Fe$_2$) | 9,337 | 910–2,000 | $6.0\times10^{-2}$ | 550 Oe (3,022 MHz)* | 0.301 - 0.662 | [4, 12, 48] |
| Fe$_{1-x}$Ga$_x$B$_y$ ($x = 0 - 0.21; y = 0.9 - 0.17$) | 12,000 | 70 | $1.0\times10^{-3}$ | 20 Oe (128.8 MHz)* | 0.543 | [4, 49] |
| (Fe$_{90}$Co$_{10}$)$_{78}$Si$_{12}$B$_{10}$ | 16,900 | 158 | $0.8–1.5\times10^{-2}$ | 40 Oe (293 MHz)* | 0.539 | [50, 51, 52, 53] |
| Galfenol (Fe$_{1-x}$Ga$_x$ ($0.12 \leq x \leq 0.3$)) | > 20,100 | 300–400 | $1.2-6.0\times10^{-3}$ | 225 Oe (1,916 MHz)* | 0.157 - 0.209 | [4, 54, 55] |

**Table 1:** Extracted parameters from V[TCNE]$_x$ strain devices compared to YIG, Terfenol-D, and other magnetostrictive materials. Asterisk indicates the frequency-equivalent linewidth calculated from the field-swept FMR linewidth and accounts for the ellipticity of FMR precession for in-plane magnetized materials following the method in Ref. [56].

# Supplemental Information: *In situ* electric-field control of ferromagnetic resonance in the low-loss organic-based ferrimagnet V[TCNE]$_{x\sim 2}$

**Cavity X-band FMR of Sample 2**

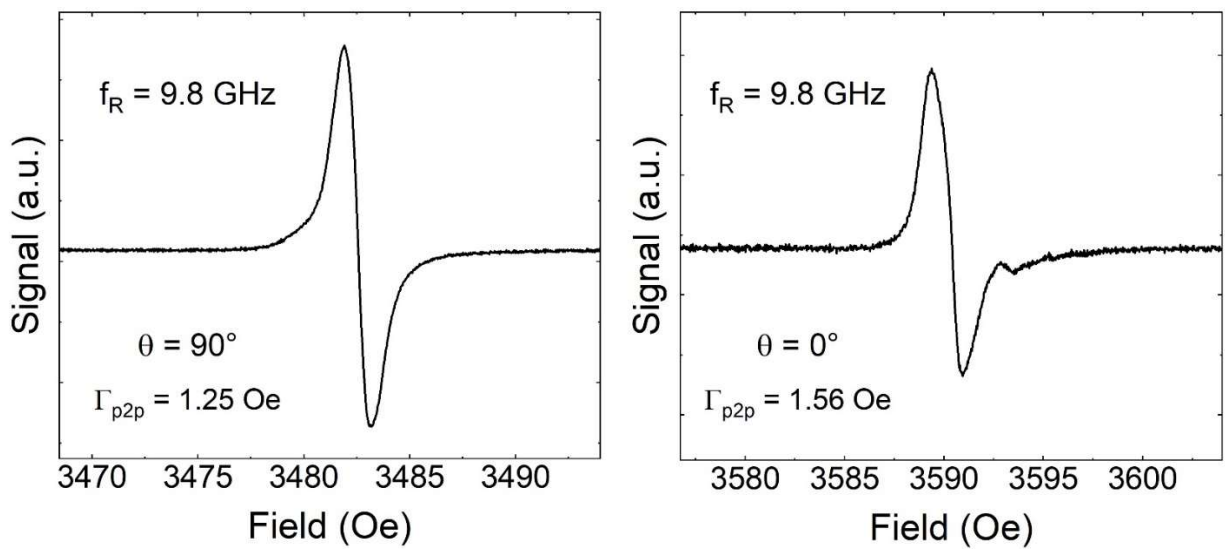

**Figure S1:** X-band (9.8 GHz) cavity FMR scans of Sample 2 for DC magnetic field in-plane ($\theta = 90°$) and out-of-plane ($\theta = 0°$). Peak-to-peak (p2p) linewidths and resonance positions determined from a fit to a Lorentzian derivative, from which the full-width-half-max (FWHM) linewidth $\Gamma$ is found by multiplying by $\sqrt{3}$.

# V[TCNE]$_x$ Resonance Frequency with Piezo Switching Strain

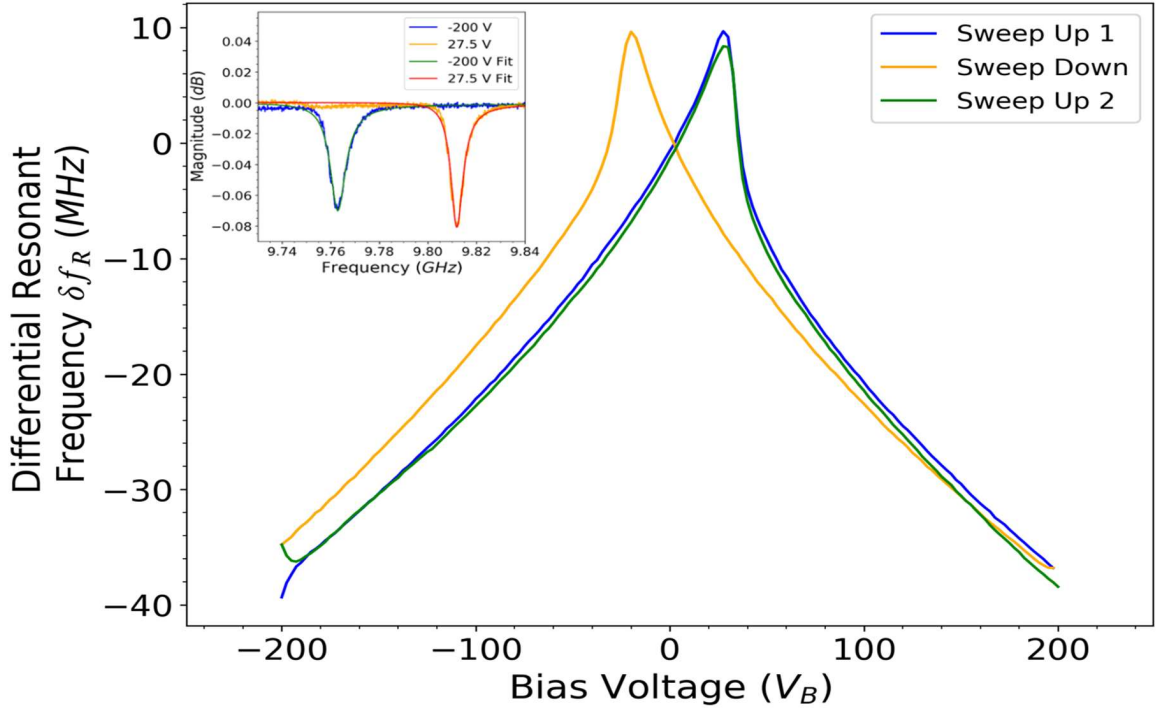

**Figure S2:** V[TCNE]$_x$ out-of-plane magnetized differential resonance frequency $\delta f_R = 9.8\ GHz - f_R$ as a function of applied bias voltage to 150 μm PMN-PT. The "butterfly" hysteresis arises from the hysteretic behavior of the piezo strain upon the polarization direction switching. The inset shows frequency-swept FMR spectra and fits at maximum and minimum frequency shift.

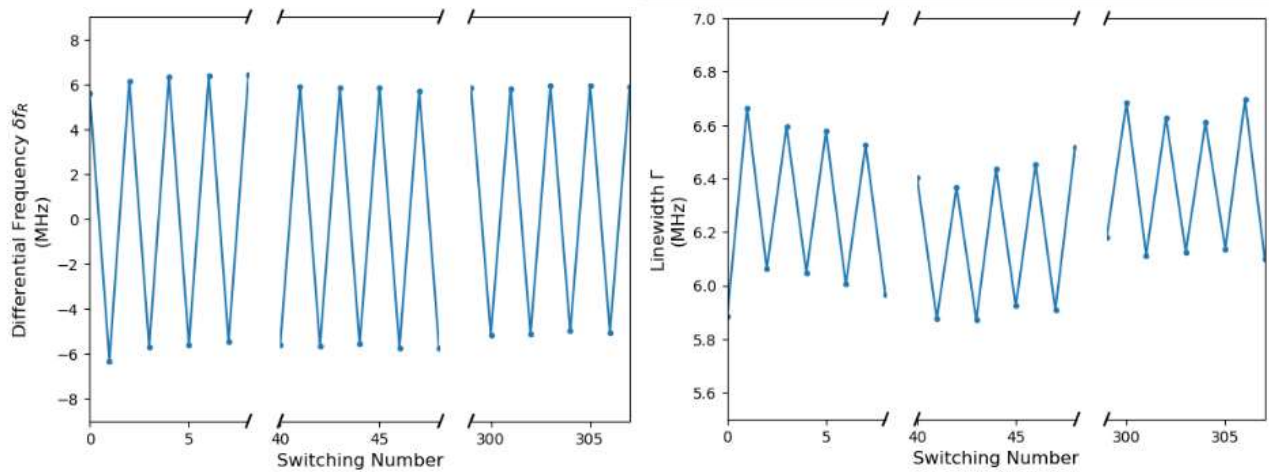

**Figure S3:** Differential resonance frequency ($f_{R,0} = 9.8$ GHz) in the same device from Fig. S2 switching between $V_B = \pm 25$ V ($E_B = \pm 1.67$ kV/cm) as a function of the number of the number of switches between positive and negative applied strain. The FWHM linewidth of the V[TCNE]$_x$ remains effectively constant for over 300 positive/negative (tensile/compressive) strain applications, and the resonant frequency for the respective compressive and tensile additionally remains effectively constant. These conditions were selected based on the linewidth and resonance frequency tuning such that the resonance features do not overlap, thereby demonstrating a means to electrically-bias a device on and off resonance.

# V[TCNE]$_x$ Density Functional Theory Calculations of Strain-Dependent Magnetoelastic Energy

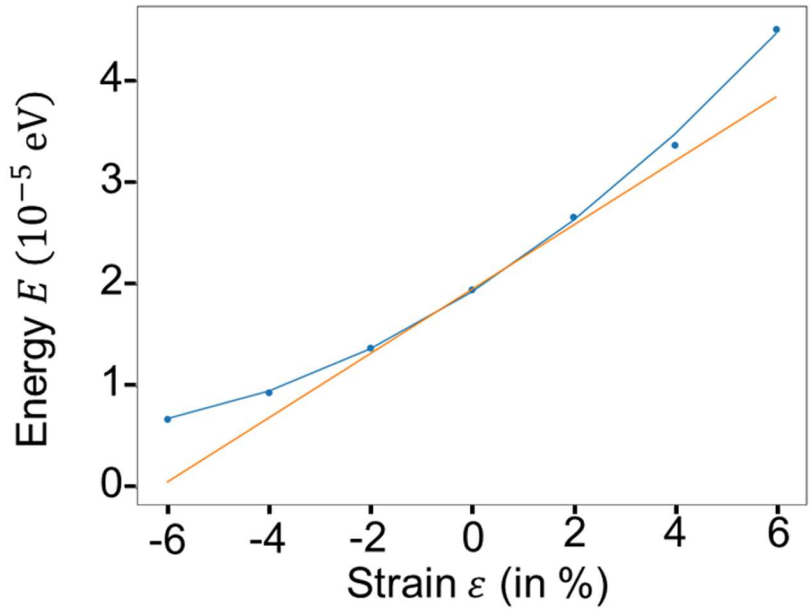

**Figure S4:** DFT-calculated magnetic energy difference of the V[TCNE]$_x$ unit cell upon manipulating the applied strain. The orange line is a tangential linear fit at $\varepsilon = 0$ to solve for $\Delta E/V$ in the main text that provides the magnetoelastic coupling $B_1$.

# V[TCNE]$_x$ Optical Strain Characterization

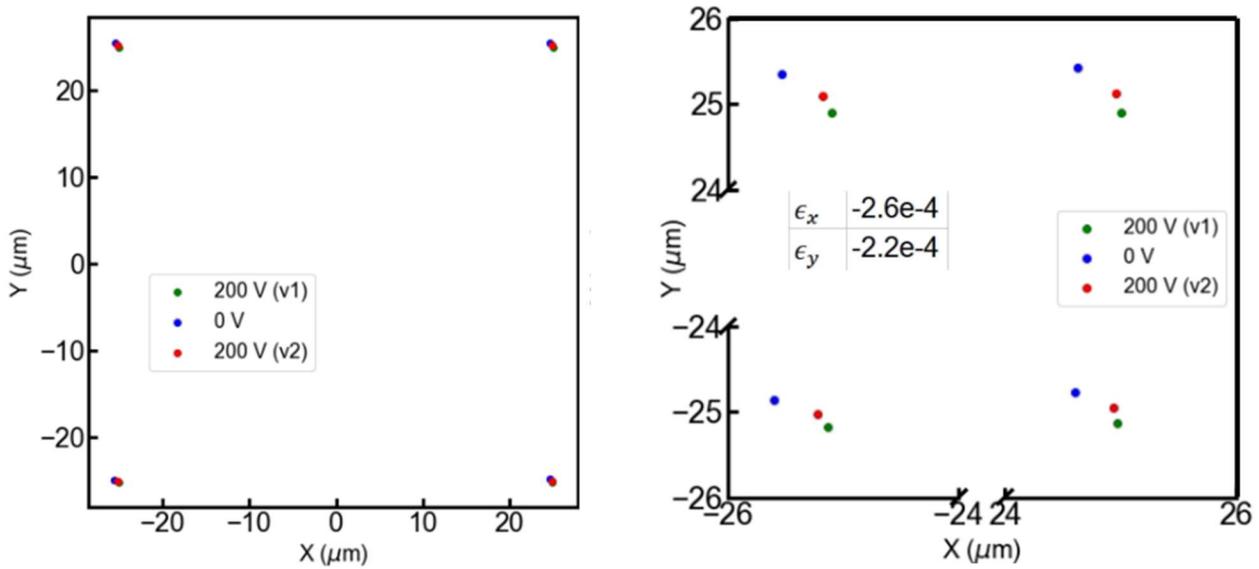

**Figure S5:** Positions of fiducial marks "burned" onto the V[TCNE]$_x$ film measured via optical techniques upon biasing a 150 μm PMN-PT piezo transducer. The extracted strain in $x$ and $y$ is averaged to $\varepsilon = 2.4 \times 10^{-4}$ and is used to calculate the magnetoelastic coefficients $\lambda_S$ presented in the text.

# Additional V[TCNE]$_x$ Device Strain Characterizations

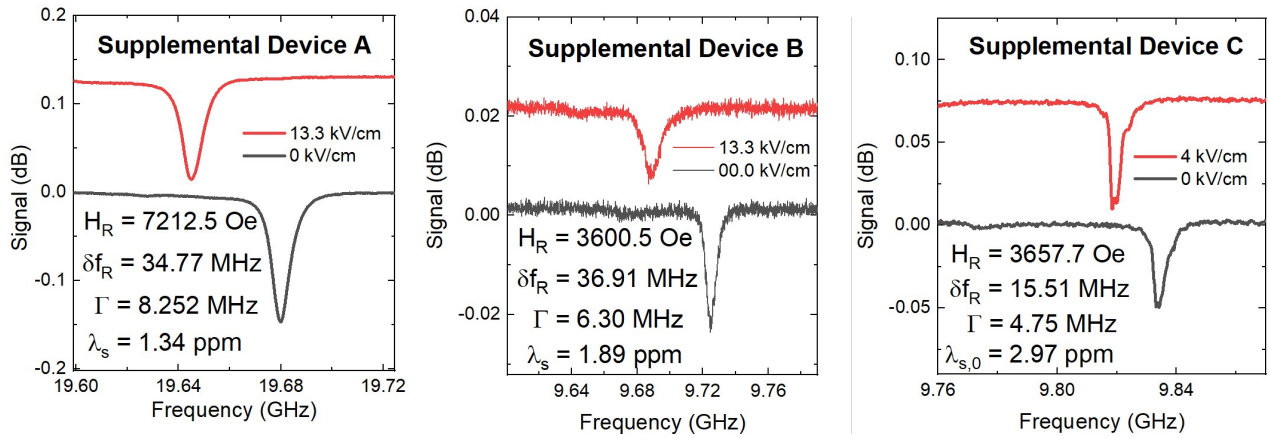

**Figure S6:** Supplemental devices measured via BFMR techniques ($\theta = 0°$). Supplemental device C varies from the others only by the thickness of the PMN-PT piezo ($t_P = 500\ \mu m$), so that the electric field across the device (hence the strain) is adjusted accordingly.

# Additional Linewidth and Damping Analysis: Supplemental Device C ($t_P = 500\ \mu m$)

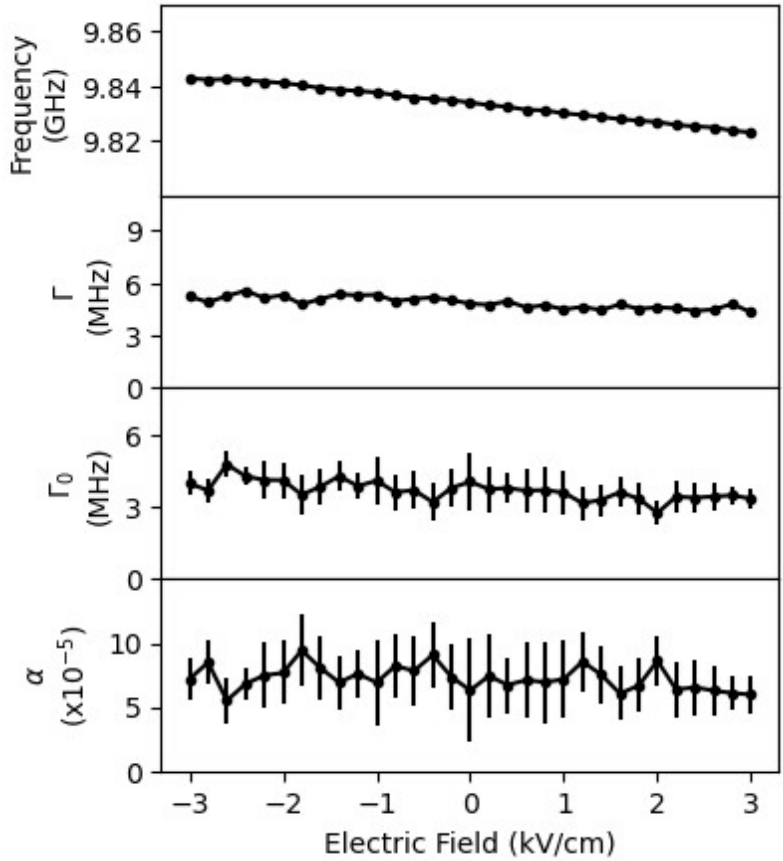

**Figure S7:** Gilbert analysis of Supplemental Device C as a function of applied electric-field bias. (a) Resonance frequency for an out-of-plane magnetization orientation and applied external field $H_R = 3{,}658.8$ G ($f_{R,0} = 9.83$ GHz). (b) FWHM linewidth corresponding to the resonance frequencies in panel (a). (c) Inhomogeneous broadening and (d) Gilbert damping parameters. Note there is negligible change in linewidth, inhomogeneous broadening, and Gilbert damping for positive *and* negative bias up to the piezo switching fields at $\pm 3.2$ kV/cm. The error bars in (a) and (b) are smaller than the markers used for the data points.